\documentclass{IOS-Book-Article}     
\usepackage{graphicx}
\usepackage{tikz-cd}
\usepackage{url}

\begin{document}
\begin{frontmatter}          
%
\title{Ontologising Trustworthy in the Telecommunications Domain}
\runningtitle{Ontologising Trustworthy in the Telecommunications Domain}

%
\author[A]{\fnms{Ian} \snm{Oliver}},
\author[B]{\fnms{Pekka} \snm{Kuure}}
\author[C]{\fnms{Wiktor} \snm{Sedkowski}} 
\author[D]{\fnms{Thore} \snm{Sommer}}
\runningauthor{}
\address[A]{Nokia Bell-Labs, Espoo, Finland\\ian.oliver@nokia-bell-labs.com\\\url{https://orcid.org/0000-0002-8319-2612} }
\address[B]{Nokia Mobile Networks, Espoo, Finland}
\address[C]{Nokia Strategy and Technology,  Wroc\l{}aw, Poland}
\address[D]{Christian-Abrecht-Universit\"at zu Kiel, Germany}

\begin{abstract}
Based upon trusted and confidential computing platforms, telecommunications systems must provide guaranteed security for the processes and data running atop them. This in turn requires us to provide trustworthy systems. The term trustworthy is poorly defined with corresponding misunderstanding and misapplication. We present a definition of this term, as well as others, demonstrate its application against certain telecommunications use cases and address how the learnings from ontologising these structures contribute to standardisation and the necessity for FAIR ontologies across telecommunications standards and hosting organisations.
\end{abstract}


\end{frontmatter}


\section*{Introduction}
The telecommunications domain is build on an extensive set of interoperable standards mediated by ETSI, GSMA, 3GPP, IETF etc, covering almost every aspect from radio physics to cloud and service architectures. This process, described by some \cite{parssinen2002pattern} as Machiavellian or as a prisoner's dilemma with multiple, sometimes irrational opponents, generates specifications that have enabled to the present day, five generations of telecommunications systems spanning over 40 years to the current day work on sixth generation (6G) standards.

Standards adhere to the `FRAND' or fair, reasonable, and non-discriminatory terms of usage \cite{contreras2015brief} and terminological agreement must be made, not just at a technical level but also at the legal and patenting levels. The use of terminologies and concepts adhering to findable, accessible, interoperable and reusable (FAIR) principles in therefore critical; and more so when standards cross multiple standardisation bodies and are integrated and referenced to. Such terminologies or ontologies are found inside the standards documents and rarely are presented as their own stand-alone terminological or ontological documents. This causes problems when terminology is then inconsistently applied across standards, even in the same series.

The use of the term \textit{trustworthy} is becoming endemic within standardisation and is especially used in telecommunications because of the critical nature of the systems and the critical nature of data being processed \cite{brusilovsky20205g}. Increasing applied to almost everything, for example \textit{trustworthy} artificial intelligence in telecoms systems \cite{kamaruddin2023compliance} is and will be a focus of 6G. The term \textit{trustworthy} is not defined and nor is it rooted or linked in any other definition of the term, and associated terms in any other standard, especially those specifically dealing with trust.

The most interesting specifications are the IETF Remote Attestation Procedures (RATS) and IETF Trusted Execution Environment Provisioning (TEEP) which both define a number of terms related to trusted and confidential computing \cite{tschofenig2019cyberphysical}. These are directed at the supporting architectures (and may be included in the later architectural models here) but do not explicitly address the definition of \textit{trustworthy}.

The term \textit{trustworthy} originates from the concepts of trusted computing and the later confidential computing - mechanisms for establishing the identity, integrity of systems, and then ensuring that any workload (e.g.: AI processes, telecoms databases etc) that requires protection (privacy, secure processing) can run in a secure environment: a trusted and attested machine \cite{Turcanu2021,risto2023forensics,borgercloud}, a CPU enclave etc. \cite{jha2022trusted}.

A trustworthy system is the minimal basis for providing any form of confidential computing \cite{mulligan2021confidential,sardar2023confidential}. As this is the chosen technology for providing `trustworthy AI' and securing data processing, ensuring a good, common understanding has wide reaching effects.

This introduces a further problem of \textit{how} trust is established \cite{birkholz2021remote}, how that trust is transferred and limited across the system - the chain of trust - and how that trust is maintained over time. 

In this paper we present the outline ontology for trusted and confidential computing, the underlying semantics and its use in selected telecoms domains and how the terms related to trust can be utilised. From this we can properly frame the discussion about what makes something \textit{trustworthy} and if presented as its own standard or supporting documentation can be utilised as a common references. We conclude with our learnings and how this work proceeds with regards to standardisation and their implementations.

\section*{Semantics of Trustworthy}
The definition of trustworthy according to the Cambridge English and Merriam Webster dictionaries:
\begin{quote}
\textbf{trustworthy}, adjective. \textit{Deserving of trust, or able to be trusted}, \textit{dependable or worthy of confidence}
\end{quote}

Trusted Computing based on some root of trust with necessary infrastructure such as Hardware Security Module (HSM) or Trusted Platform Module (TPM), provides a mechanism for establishing and endorsing trust. Remote attestation \cite{oliver2021trust} as part of the management, operations, and supply-chain provides verification point.\footnote{The Nokia Attestation Engine is one implementation of this: \url{https://github.com/nokia/AttestationEngine}}.

\begin{figure}[ht]
\begin{center}
\begin{tikzcd}
Element \arrow[r, "measure"] \arrow[rr, "attestable"', dashed, bend right] \arrow[rrr, "trustworthy", dashed, bend right=49] & Measure \arrow[r, "attest"] & Claim \arrow[r, "verify"] & Result \arrow[r, "decide"] & Decision
\end{tikzcd}
\end{center}
\caption{Definitions of Trust}
\label{deftrust}
\end{figure}
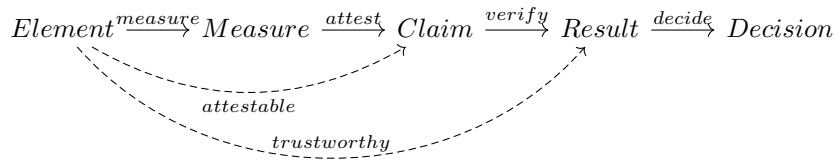

Remote attestation is based around the idea of measurement, for example, cryptographic hashes calculated during the boot-time of a device, or as part of file system integrity monitoring. This further involves collecting those measurements - known as claims - and verifying them against known good values. Once this is complete, a decision is made whether to trust the device. This gives four actions: measure, attest, verify and decide as expressed in figure \ref{deftrust} utilising  the formalisms here \cite{oliver2009visualizing,cheng2022joy}.

If measurements - cryptographic hashes, boot logs, identity structures, TPM/SGX quote structures etc. - can be taken and the element has some interface providing access to those then the element is \textit{attestable} and a claim produced. Good examples of claims are the TPM 2.0 Quote and SGX tcbblock structures, while the IEFT RATS Entity Attestation Token (EAT) \cite{ballesteros2020rats} is a structure for generalising this. 

If the correctness of the obtained claim can be verified then the element is \textit{trustworthy} - note the dictionary definition above. That this does not mean we trust the element, but rather that we can obtain information that we can utilise to make a confident decision whether it is trusted or not.

\begin{figure}[ht]
\begin{center}
\begin{tikzcd}
TrustedElement \arrow[r] \arrow[dd, tail]     & 1 \arrow[dd, "true" description] \\
                                              &                                  \\
Element \arrow[r, "\phi_{Decision}"', dashed] & Decision                        
\end{tikzcd}
\end{center}
\caption{Trusted Elements}
\label{trustedelements}
\end{figure}
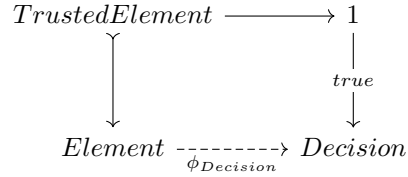

Figure \ref{trustedelements} shows how the notion of trust\footnote{As a categorical sub-object classifier} is formed. Trusted elements are those elements (a subset of in this formalism) those elements that map to positive decisions. The statement $\phi_{Decision}$ represents the whole measurement, attestation, verification and decision process. Decisions are not necessary binary true/false in nature and from this we could\footnote{and in the Nokia Attestation Engine we do} have intermediate values, to handle explicitly network failure, incorrect requests etc, which in turn provides the possibility of degrees or levels of trust and their [partial] ordering. This is out of scope for this discussion.

The above diagrams (figs.\ref{deftrust} and \ref{trustedelements}) are simplified - we do not show the mechanisms for internal structures and their composition\footnote{Objects, such as elements, measures etc, are composed utilising push-outs which very naturally gives us structures such as the TPM2.0 quote.} - but forms the core principles for these terms, how they are formed and what they mean.

This gives our semantic framework for trust and trustworthiness and relates the objects in our system: elements, measure, claims, results and decisions, with the actions over these and the descriptions such as trusted, attestable, trustworthy \cite{oliver2019exploring}. In the following sections we apply this to the concepts and models found in telecommunications systems.

\section*{NFV, O-RAN and 5G}
\begin{figure}[ht]
        \centering
        \includegraphics[scale=0.34]{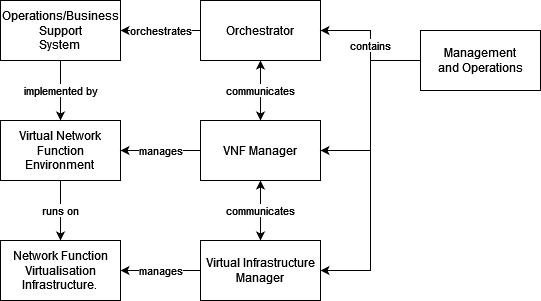}
        \caption{Network Function Virtualisation Elements}\label{nfv}
    \end{figure}

\begin{figure}[ht]
        \centering
        \includegraphics[scale=0.34]{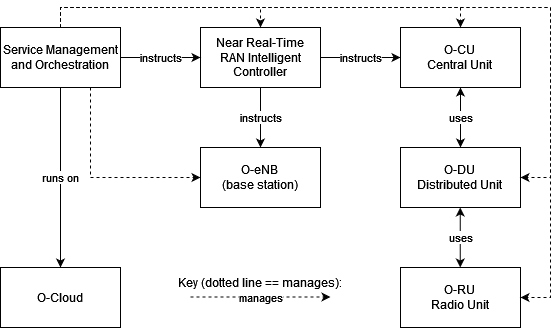}
        \caption{Open Radio Access Network Elements}\label{oran}
    \end{figure}

\begin{figure}[ht]
        \centering
        \includegraphics[scale=0.34]{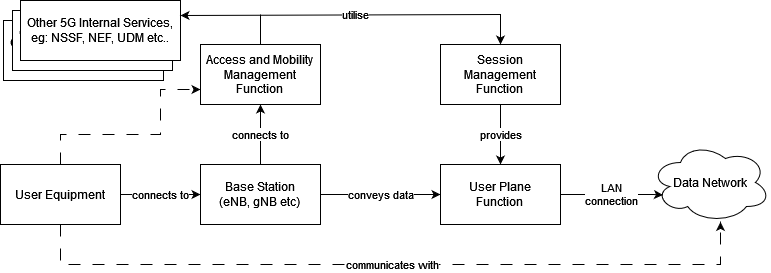}
        \caption{5G System Elements}\label{5G}
    \end{figure}

The ETSI Network Function Virtualisation Architecture \cite{ersue2013etsi} and Open Radio-Access Network Architecture \cite{singh2020evolution} are two of the main components of modern telecommunication systems \cite{wiley5gref}. These each have their own architecture models which serve to define the major functional components and their interfaces. These are used to define terminology and plan the standardisation processes, especially in terms of the functionality and data required by the interfaces. The elements in these models should not be taken to be component definitions but rather than broad roles the implementing components take; in the highly simplified (and incomplete!) figures \ref{nfv}, \ref{oran} and \ref{5G} we utilise the relationships between components rather than their interface names as from the specification \cite{oliver2018modelling}.

\begin{itemize}
    \item O-Cloud is constructed from [physical] infrastructures nodes. This may conform to the NFV architecture principles, and some may reside within a 5G deployment.
    \item O-eNB is an eNB (4G) or gNB(5G) that supports the E2 (SMO-*NB) interface
    \item A 5G base station (gNB) comprises of from an O-RU, O-DU and O-CU and potential additional services.
    \item O-CU, O-DU and the 5G services (AMF, SMF, UPF etc) can be realised and deployed as bare-metal processes or as virtualised functions, typically containers but also as virtual machines.
    \item O-RU are physical network function with one or more antennas. O-DU may be co-located with the O-RU for real-time requirements and constraints
    \item An O-RU, being a physical computing (albeit specialised) unit may managed as part of the NFVI
    \item O-DU and some 5G services may be distributed towards the network edge.
    \item Customer services provided OSS/BSS may be implemented as bare-metal, containers etc., and distributed accordingly.
    \item O-RAN SMO and NFV MANO will be implemented similarly to other services.
\end{itemize}

\begin{figure}[ht]
        \centering
        \includegraphics[scale=0.34]{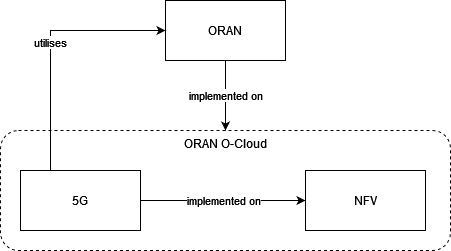}
        \caption{5G, O-RAN, NFV Relationships}\label{5gorannfv}
\end{figure}

We can further describe these relationships by relating the implementation to the network function virtualisation infrastructure (NFVI in fig.\ref{nfv}) which defines a collection of computing elements - in effect, things with CPUs and in our case trusted elements such as TPMs, CPU enclaves etc. It is important to note that that radio units, despite their internal complexity including ASICs and other specialised hardware for processing radio signals are in effect just computers as any other.

\begin{figure}[ht]
        \centering
        \includegraphics[scale=0.34]{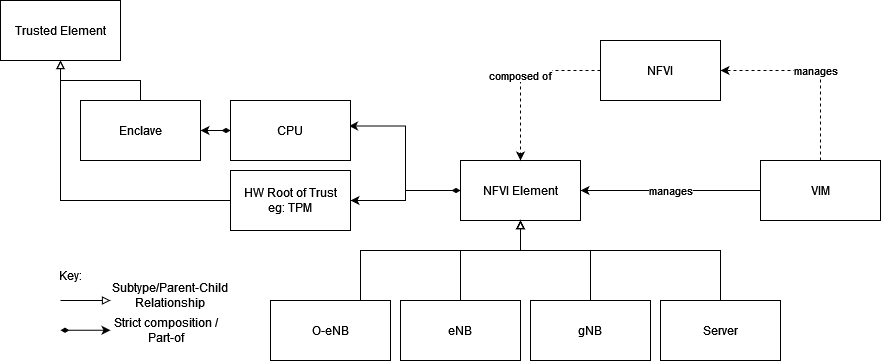}
        \caption{Implementation via NFVI Elements}\label{implementation}
\end{figure}

\section*{Providing Trust}
\begin{figure}[ht]
        \centering
        \includegraphics[scale=0.34]{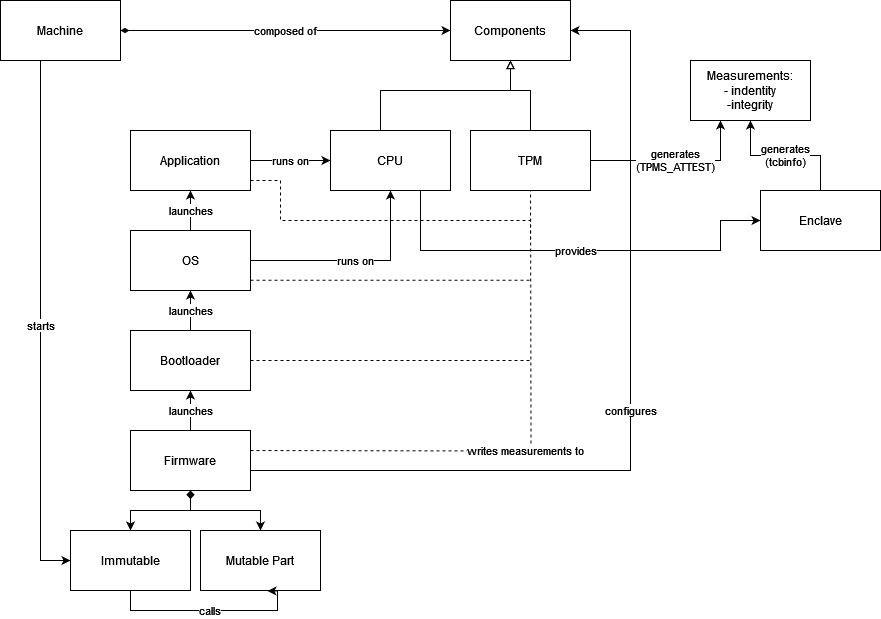}
        \caption{Trust Elements}\label{trustthings}
\end{figure}

As discussed earlier trust in a system is provided from a core root-of-trust using hardware root-of-trust modules such as the TPM or some other mechanism such as that found in CPU enclaving systems such as Intel SGX, AMD SEV-SNP and TrustZone. Figure \ref{trustthings} describes these elements.

A device consists of a number of components which are capable of running network applications. Device power-on from immutable code \cite{tan2017review}; the configuration will vary between ASICs, FPGAs, PCs, microcontrollers up to mainframes and supercomputers.

The measured boot process \cite{tomlinson2017introduction} in forms such as the UEFI measured boot on x86 devices, Dynamtic Root-of-Trust found in Intel TXT equipped systems, similar in some Power based systems and [very] occasionally in IoT devices \cite{zaidenberg2018hardware}, starts with an explicitly trusted piece of code \cite{tcgefi}, which then cryptographically measures (hashes) the next piece of code to run - the firmware proper - and write this to a TPM. This process continues during boot until run-time forming a chain-of-trust through a Merkel tree of hashes describing the system's integrity \cite{yao2020maintenance}.

Secure boot is a complimentary process providing verification of a component or software through digital signatures and prevents locally the execution of unverified components. Secure boot only provides evidence about the provenance and not the validity of the contents. If the signing keys are lost - an unfortunately regular occurrence - then anyone can potentially sign components.

\begin{figure}[ht]
        \centering
        \includegraphics[scale=0.34]{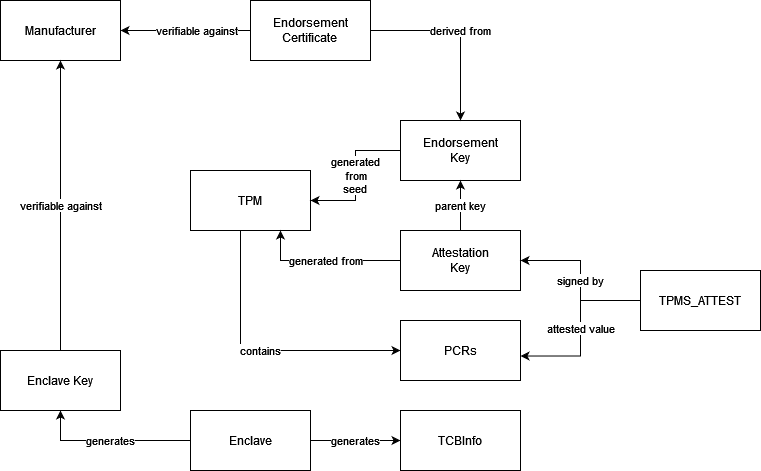}
        \caption{Quotes and Their Derivations and Relationships}\label{quote}
\end{figure}

Figure \ref{quote} shows some of the more important \textit{quote} structures which provide cryptographically signed evidence or claims about the machine's identity and integrity.

We have abstracted away a number of details, for example, a CPU enclave generating an enclave key is a very specific and complex process. However these vendor specific details are not required at this level.

Important to note is the amount of cross-referencing, such as the TPM's Endorsement Certificate being verifiable against a manufacturer's certificate authority, or the UEFI eventlog's contents used to calcualte PCRs values etc. The TPM's quote (TPMS\_ATTEST) structure itself is signed by an attestation key (AK) derived from the endorsement key (EK) and contains a hash of the hashed public parts of the AK and the EK.

\begin{figure}[ht]
\begin{center}
\begin{tikzcd}
                                                             & t:TPM \arrow[rr, "measure"]    &  & ak:Key \arrow[r, dotted]     & q:TPMQuote                          \\
e:PC \arrow[rd, dotted] \arrow[r, dotted] \arrow[ru, dotted] & f:FW \arrow[rr, "measure"]     &  & pcr0:Hash \arrow[r, dotted]  & pcr0 || pcr1:Hash \arrow[u, dotted] \\
                                                             & c:Config \arrow[rr, "measure"] &  & pcr1:Hash \arrow[ru, dotted] &                                    
\end{tikzcd}
\end{center}
\caption{Elements and Measures}
\label{elementmeasures}
\end{figure}
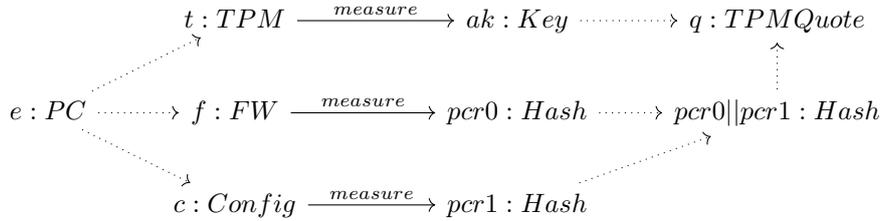

Referring back to figure \ref{deftrust} these are the measurements from any given element. Figure \ref{elementmeasures} shows how these (with composition) link together in a specific example: here a PC's firmware and firmware configuration form measurements and the composition of these a quote which becomes a claim through the process of attestation.

This shows the structure of a PC composed of a TPM, firmware and configuration, each of these with its own measurement(s). A TPM quote is a measurement formed by the composition of these measures. Composition of PCRs means an ordered concatenation and hashing (Merkel tree) and composition of a Key means signing by that key. 

It is worth noting the fields of the TPM's quote or TPMS\_ATTEST structure as described in table \ref{tpmsattest} - the quote structure for SGX can be found in \cite{intelsgxquote}.

\begin{table}[ht]
\begin{center}
\begin{tabular}{l|l}
   \textbf{Field} & \textbf{Description} \\\hline
   pcrDigest  &  hash of the selected PCR entries \\
   pcrSelect  &  list of the selected PCR fields \\
   magic, type & Fixed values denoting the type of the structure (TCG defined) \\ 
   firmwareVersion & Firwmare version of the TPM 2.0 device \\
   clockInfo::clock & Value of the TPM's internal clock \\
   clockInfo::resetCount & Counters showing the number of power cycles \\
   clockInfo::restartCount & Counters showing the number of sleep (S3) cycles \\   
   clockInfo::safe & Flag showing whether the TPM was powered off explicitly\\
   extraData & Any user supplied data, eg: a nonce for replay attack prevention \\
   qualifiedSigner & a hash of the public parts of the signing key's hierachy
\end{tabular}
\end{center}\caption{TPM 2.0 Quote (TPMS\_ATTEST) Fields}\label{tpmsattest}
\end{table}

It follows that the term `integrity of a system` is just the  composition of all possible measures for that system. This implies a lattice in which we can delineate a bound above which gives a minimal set of sufficient measurements that must be taken.

\section*{Trusting a Telecommunications Deployment}
Utilising the models in the earlier sections we apply these to some selected use cases for prototypical telecommunications deployments. These are used to both to validate our models and assist in deriving the properties a `trusted/trustworthy' O-RAN/5G/NFV system requires. It is exactly these use case that are used to drive the standardisation processes and the notions of \textit{trustworthy}.

\subsection*{O-RU Trust in NFV}
\begin{figure}[ht]
        \centering
        \includegraphics[scale=0.34]{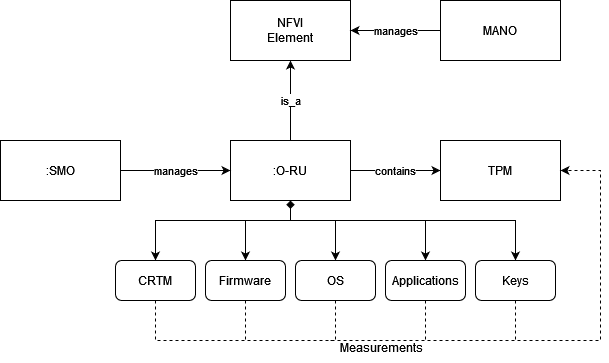}
        \caption{O-RU Relationships}\label{orutrust}
\end{figure}

A simple case is an O-RAN Radio Unit is an NFVI element containing in this case a TPM - this is a typical deployment. The O-RU provides an attestation interface towards an attestation server within the O-RAN SMO as shown in in figure \ref{orutrust}.

The following questions deriving from this must be answered in order to establish trust:
\begin{itemize}
    \item Which measurements and combinations are are required?
    \item What is the decision mechanism by which O-RAN SMO denotes the O-RU as trusted?
    \item What is the decision mechanism by which NFV MANO denotes the O-RU as trusted?
    \item What is the correlation between the O-RAN SMO view of trust and NFV MANO view of trust?
\end{itemize}

We may also establish failure scenarios such as if the NFV MANO detects a loss of trust but the SMO not, is the O-RU still trusted? Is there a trust ordering such that $x:RU\;|\;trusted_{SMO}(d) \Rightarrow trusted_{NFV}(d)$?

\subsection*{O-DU Trust}
\begin{figure}[ht]
        \centering
        \includegraphics[scale=0.34]{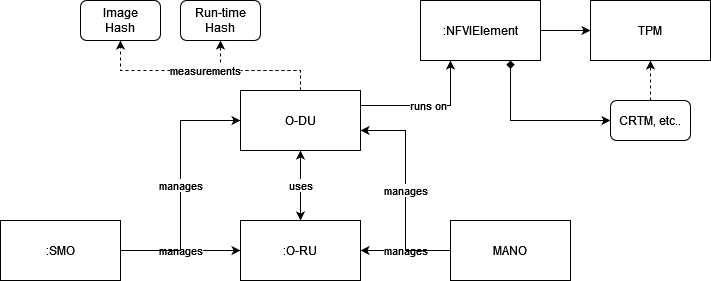}
        \caption{O-DU Relationships}\label{odutrust}
\end{figure}

The question of O-DU Trust is more complex; the DU is often deployed as a container, which in turn runs on some NFVI element. The DU uses one or more RUs as part of its operation. Figure \ref{odutrust} shows a scenario - NB: the RU here is the same RU (with measurements) as in fig.\ref{orutrust}.

Again, we infer similar questions about the NFVI Element and how MANO decides on its trust; the O-RU is as earlier. For the O-DU we have the following questions inferred:

\begin{itemize}
    \item Can an O-DU only run on trusted NFVI elements?
    \item Does a similar trust ordering exist as before: $d:DU\;|\;trusted_{SMO}(d) \Rightarrow trusted_{NFV}(d)$?? 
    \item Can an O-DU manage untrusted O-RUs and vice versa, can a trusted O-RU be utilised by an untrusted O-DU?
\end{itemize}

\subsection*{O-DU Confidentiality}
\begin{figure}[ht]
        \centering
        \includegraphics[scale=0.34]{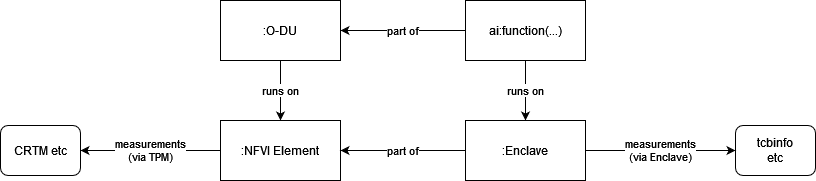}
        \caption{O-DU Confidentiality Relationships}\label{oduconfidential}
\end{figure}

The final case depicted in figure \ref{oduconfidential} involves confidential computing and that some functionality (maybe some AI function that needs to be trustworthy?) within a O-DU must run within a CPU enclave.

The questions raised earlier again are all valid here, but now with the following additions:
\begin{itemize}
    \item Does a relationship of trust exist between the enclave and the NFVI element on which that enclave is implemented (via the CPU)
    \item Does the enclave need to be on the same NFVI element? Does this imply `enclave mobility' or is there a `locality' of processing constraint?
    \item Can a function requiring an enclave run on an untrusted machine, even though the enclave may be trustworthy (and trusted)? 
    \item How is trust communicated across the composition (part-of) relationships?
\end{itemize}

In current trustworthy systems there is no direct mechanism to relate the quote generated by an enclave with the quote generated by the the host machine's TPM. If a link does exist then it is very ephemeral and hidden in firmware measurements during boot (if at all, and even then it is most likely proprietary and unavailable, which fails our attestatble definition from earlier).

From this model we are forced to address the nature of composition relationships and the establishment and maintenance chain-of-trust over these relationships. We are also forced to confront the fact that a function may be trusted, but is part of an component and running on trusted elements. Putting it in this manner it becomes obvious that \textit{trustworthy} of individual components is not the same as \textit{trustworthy} when applied to larger systems.

\section*{Conclusions}
We have presented the domain, and set out to define the underlying semantics for the terms trust, trustworthy, attestable and outlined how these terms fit together. We have then shown an example applying this to the O-RAN, 5G and NFV concepts which underpin modern telecommunications systems. Obviously for brevity much has been simplified in how these systems are composed but satisfies the definition of ontology itself, in that we provide a mechanism for the kinds of things and their nature in these systems.

Through the use of some simple examples drawn from the telecoms domain we have shown that we can easily generate a large number of questions about how trustworthiness is establish, how a trust decision might be made, as well as identifying where attestation must take place. Larger questions then arise from how the chains-of-trust and system composition interact.

From a FAIR point of view it has been obvious to many in this domain that such ontologies must be made to provide a common terminological grounding. In day-to-day work we interact with over ten international standards organisations specifically for telecommunications - this does not include governmental regulations and internal specifications and requirements.

Any such ontology in this area is most likely an evolving document rather than a fixed standard. It is also necessary that the terms here do not conflict but are mappable to existing terms. For example, the use of EAT in IETF RATS is a `claim' in our ontology, and its contents `measurements.

Implementation of the ontologies here has been made in Protege and will be released as open source, possibly via one or more standards organisations. At minimum these ontologies will remain open, freely accessible and become part of any input to standards relating to cybersecurity across the standardisation organisations.

A final note on the complexity of the ontologies: in the current model we have approximately 100 types, 200 relationships and the description logic is $\mathcal{ALCHIF(\circ,*)}$.

\bibliographystyle{vancouver}
\bibliography{refs} 


\end{document}